\documentclass[runningheads]{llncs}

\usepackage{listings}
\usepackage{algpseudocode}
\usepackage{amsmath}
\usepackage{graphicx}
\usepackage{caption}
\begin{document}

\title{A Streaming Analytics Language \\
for Processing Cyber Data}

\author{Eric L. Goodman \inst{1,2} \and Dirk Grunwald\inst{2}}

\institute{Sandia National Laboratories, Albuquerue, NM, USA \and
CU Boulder, Boulder, CO, USA} 

\maketitle

\begin{abstract}
We present a domain-specific language called SAL 
(the Streaming Analytics Language) for processing
data in a semi-streaming model.  In particular we 
examine the use case of processing netflow data 
in order to identify malicious actors within a network.  
Because of the large volume of data generated from
networks, it is often only feasible to process the data 
with a single pass, utilizing a streaming
($O(polylog \, n)$ space requirements) or semi-streaming 
computing model ( $O(n \cdot polylog \, n)$ space requirements).  
Despite these constraints, we are able to achieve an average of 
0.87 for the AUC of the ROC curve for a set of situations 
dealing with botnet detection.  The implementation of an
interpreter for SAL,
which we call SAM (Streaming Analytics Machine), 
achieves scaling results that show
improved throughput to 61 nodes (976 cores), 
with an overall rate of 
373,000 netflows per second or 32.2 billion
per day.  SAL provides a succinct way to describe 
common analyses that allow cyber analysts
to find data of interest, and SAM is a scalable interpreter of the language.
\end{abstract}

\section{Introduction}

Cyber security is challenging problem due to the large volume of
data that is produced, the changing nature of the data, and
the ever evolving threat landscape.  Cyber analysts need
to extract features from a high-throughput stream, create
models that will predict malicious behavior or anomalies,
evaluate the results, and iterate in a continuous cycle of
improvement and adjustment.  Our work provides a domain
specify language (DSL) for expressing streaming
computations on cyber data, enabling cyber analysts
to quickly express machine learning pipelines to analyze and
classify the data.
The main contribution of this paper is to combine within one DSL
the ability to succinctly express 
streaming operators,
vertex-centric graph operations, and
machine learning pipelines.
%\vspace{-0.75\topsep}
%\begin{itemize}
%\itemsep-.25em
%\item streaming operators,
%\item vertex-centric graph operations, and
%\item machine learning pipelines.
%\end{itemize}
%\vspace{-0.5\topsep}
We call this language the Streaming Analytics Language (SAL).
While this work focuses on detecting malicious activity
in high-volume cyber data, SAL can be applied to any streaming problem 
where the elements of the stream are tuples.  

We extract features from a stream
of cyber data, where the data generation rate only allows a single pass.  
Streaming \cite{Muthukrishnan:2005:DSA:1166409.1166410} and semi-streaming 
\cite{Muthukrishnan:2005:DSA:1166409.1166410,Feigenbaum:2005:GPS:1132633.1132638} fit 
the requirements.  However, it is currently
cumbersome to express streaming operators as no 
high-level language currently has them
as first-class citizens.  
%Additionally, many desirable features include a
%temporal dimension.  SAL allows users to express temporal quantities about incoming netflows.
Also, it is often desirable to extract queries in terms of nodes within the graph of network activity.
For example, we may want to gather statistics about the incoming flow sizes of individual IPs.
SAL makes this type of vertex-centric computation easy to express.
Once we have features, we can define a machine learning pipeline to conduct 
either supervised or unsupervised machine learning.
      
Besides the language itself as a contribution, we also present a 
scalable implementation that translates SAL into C++ code that runs in 
parallel on a cluster of machines.  We call this interpreter the Streaming
Analytics Machine, or SAM.  We show scaling to 61 nodes on the
real-world problem of identifying malicious traffic from botnets.  
The example pipeline we describe in this paper can process over 373,000 
netflows per second, or about 32.2 billion per day. We apply the 
pipeline to CTU-13 \cite{GARCIA2014100}, which contains 13 botnet scenarios.  
Classifying at a per-netflow basis, across this dataset we achieve an 
average area under the curve (AUC) of the receiver operating characteristic 
curve (ROC) of 0.87 with the median being 0.90.  We envision SAL 
employed as a filter, so 
that analysts can concentrate more expensive analysis on a much reduced set.   

In Section \ref{section:language} we present SAL and also walk through a 
case-study of how SAL can be used to express a pipeline described in 
another paper \cite{Bilge:2012:DDB:2420950.2420969}.  In Section 
\ref{section:implementation} we describe SAM, the implementation we 
developed to interpret SAL.  Section \ref{section:classifier_results} discusses results
of detecting malicious activity and Section \ref{section:scaling_results} presents
the scaling achieved by the system.  Section \ref{section:related} compares our
work to related efforts, and Section \ref{section:conclusions} concludes.

\section{A Language for Streaming and Semi-Streaming Operations}
\label{section:language}

In this section we discuss streaming and semi-streaming algorithms and 
define how the Streaming Analytics Language expresses those algorithms.  

Streaming algorithms is a research area where
space and temporal requirements are polylogarithmic 
(i.e. $O((\log n)^k)$ for some $k$) 
\cite{Muthukrishnan:2005:DSA:1166409.1166410}.
Sometimes those constraints, in particular the temporal complexity, are relaxed \cite{PaloAltoReport1998}.  Several algorithms have been published within 
these constraints:
K-medians \cite{Babcock:2003:MVK:773153.773176},
frequent items \cite{Golab:2003:IFI:948205.948227},
mean/frequency counts \cite{Manku:2002:AFC:1287369.1287400,Datar:2002:MSS:545381.545466,Zhu:2002:SSM:1287369.1287401},
quantiles \cite{Arasu:2004:ACQ:1055558.1055598},
rarity \cite{Datar2002},
variance \cite{Babcock:2003:MVK:773153.773176,Zhu:2002:SSM:1287369.1287401},
vector norms \cite{Datar:2002:MSS:545381.545466},
similarity \cite{Datar2002}, and
count distinct elements \cite{2008arXiv0801.3552C,Metwally:2008:WGL:1353343.1353418}.
%All of the algorithms we employ in SAL require a single pass over the data and have 
%polylogarithmic space requirements.
%\begin{itemize}
%\item K-medians \cite{Babcock:2003:MVK:773153.773176}
%\item Frequent Items \cite{Golab:2003:IFI:948205.948227}
%\item Mean/Frequency Counts \cite{Manku:2002:AFC:1287369.1287400, 
%Datar:2002:MSS:545381.545466, Zhu:2002:SSM:1287369.1287401}
%\item Quantiles \cite{Arasu:2004:ACQ:1055558.1055598}
%\item Rarity \cite{Datar2002}
%\item Variance \cite{Babcock:2003:MVK:773153.773176, Zhu:2002:SSM:1287369.1287401}
%\item Vector norms \cite{Datar:2002:MSS:545381.545466}
%\item Similarity \cite{Datar2002}
%\item Count distinct elements \cite{2008arXiv0801.3552C, Metwally:2008:WGL:1353343.1353418}
%\end{itemize}

While some streaming algorithms operate on the entire stream, we focus on the
sliding window model, where only recent inputs contribute to 
feature calculation.  We believe 
the sliding window model is more appropriate for cyber data, 
where the environment is constantly changing.  
The sliding window model can also be subcategorized
into either a window over a time duration, or by the 
last $n$ items.  Currently SAL only supports expressing windows 
over the last $n$ items, but many of the underlying algorithms
are easily adapted to temporal windows, so support would
be easy to add.

Many SAL programs have $O(N \cdot polylog \, n)$ spatial requirements, where 
$N$ is the number of vertices in the graph (e.g. IPs) and $n$ is the size of the sliding window.
Generally each vertex undergoes a set of polylogarithmic operations.
This is similar to 
semi-streaming \cite{Muthukrishnan:2005:DSA:1166409.1166410,Feigenbaum:2005:GPS:1132633.1132638} 
which has  
$O(n \cdot polylog \, n)$ spatial requirements for graph algorithms where $n$ is the number
of vertices.  
%Many graph algorithms cannot be calculated on anything less then
%$O(n \cdot polylog \, n)$ space, hence the need to relax the requirement.

SAL is an imperative language.  Below is a short example.

\begin{lstlisting}[caption=SAL code example,label={lst:example},escapechar=|]
//Preamble Statements
WindowSize = 1000;  |\label{example:preamble}|

// Connection Statements
Netflows = VastStream("localhost", 9999);|\label{example:connection}|

// Partition Statements
PARTITION Netflows By SourceIp, DestIp;|\label{example:partition}|
HASH SourceIp WITH IpHashFunction;
HASH DestIp WITH IpHashFunction; 

// Pipeline Statements
VertsByDest = STREAM Netflows BY DestIp;|\label{example:stream}|
Feature1 = FOREACH VertsByDest GENERATE ave(SrcTotalBytes);|\label{example:feature}|
Filtered = FILTER VertsByDest BY Feature1 > 1000|\label{example:filter}|		
\end{lstlisting}

\begin{figure}
\centering
\includegraphics[width=.6\linewidth]{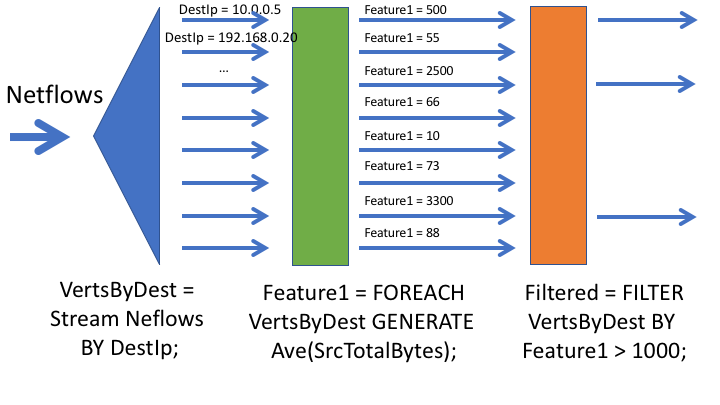}
\caption{Visual representation of the SAL program in Listing 1.  
%There is an \emph{imux} 
%operation expressed by the
%\emph{Stream $<$StreamName$>$ By $<$Key1, Key2, ...$>$} operator.  In this instance the key
%is the destination IP address, so the stream
%of netflows is converted into multiple streams partitioned by the destination IP address.
%Then a feature is created, which is the average total bytes coming from the source.
%Finally a filter is applied which down-selects, only allowing through netflows 
%where the average number of bytes coming from source IPs is greater than 1000 bytes.   
}
\label{figure:imux}
\end{figure}

Each SAL program has four parts:
preamble,
partition,
connection, and
pipeline statements.
%\begin{itemize}
%\item optional preamble statements
%\item partition statements
%\item connection Statements
%\item pipeline statements
%\end{itemize} 
Preamble statements allow for global constants to be defined that are used throughout the program.
In the above listing, line \ref{example:preamble} defines the default 
window size, i.e. the number of items in the sliding window.
%Other global constants that can be defined in the preamble are outlined in 
%Appendix \ref{appendix:parameters}.

After the preamble are the connection statements.
Line \ref{example:connection} defines a stream of 
netflows called \emph{Netflows}.  \emph{VastStream} tells
the SAL interpreter to expect netflow data of a particular format (we use the same format
for netflows as found in the VAST Challenge 
2013: Mini-Challenge 3 dataset \cite{VAST}).  Each
participating node in the cluster receive netflows 
over a socket on port 9999.  The \emph{VastStream} function 
creates a stream of tuples that represent netflows.
For the tuples generated by \emph{VastStream}, keywords are defined to 
access the individual fields of the tuple.

There are several different standard netflow formats.  SAL 
currently supports one; however, adding other
netflow formats is straightforward.  You define a C++
std::tuple with the required fields and a function object that accepts a string and returns
an std::tuple.  Once a mapping is defined from the 
desired keyword (e.g. \emph{VastStream})
to the std::tuple, this new tuple type can be used in 
SAL connection statements.  The mapping
is defined via the Scala Parser Combinator \cite{scala-parser-combinator}. 
%and is discussed in more detail in Section \ref{section:implementation}.
%In Section \ref{section:caseStudy}, we also present a mechanism for
%defining tuples in an ad-hoc fashion within a SAL program.  
  
Following the connection statements is the definition of how 
the tuples are partitioned across the cluster.  Line \ref{example:partition} specifies
that the netflows should be partitioned separately by SourceIp and DestIp.  
Each node in the cluster acts as an independent sensor and 
receives a separate stream of netflows. These independent streams 
are then re-partitioned across the cluster.  In this example, each node 
is assigned a set of Source IP's and Destination IP's using a common hash function.  
%How the IP addresses are assigned is using a common hash function.
%The hash function used is specified on lines 8 and 9, called the \emph{IpHashFunction}.
%This is another avenue for extending SAL.  
Hash functions can be defined
and mapped to SAL constructs, similar to how other tuples can be added to SAL.
The process is to define a function object that accepts the tuple type and returns
an integer, and then map the function object to a keyword using the Scala
Parser Combinator.

The last part of a SAL program is the pipeline statements which describe how the 
data is to be processed.
Line \ref{example:stream} logically separates the single Netflow stream into 
multiple streams using \emph{STREAM $<$StreamName$>$ BY $<$Key1, 
...$>$}.  In the above example,
netflows are separated into streams by the destination IP address, as can be seen in Figure
\ref{figure:imux}.  Only fields that were defined in 
the partition statement can be used in the \emph{BY} clause.

Separating out the netflows by a given set of keys allows the collection of features based on those
keys.  Line \ref{example:feature} of Listing 1 demonstrates creating a feature on the separated streams.  
For each destination IP address $x$, an average is computed on the total bytes sent to $x$ from any IP in
the sliding window.  
In general, the format for the \emph{FOREACH} statement is the following:
\lstset
{
	numbers=none
}
\begin{lstlisting}
<FeatureName> = FOREACH <StreamName> GENERATE <Operator> 
\end{lstlisting}
%[caption=FOREACH GENERATE format,label={lst:foreachgenerate}]
The available operators are \emph{ave}, \emph{sum}, \emph{topk}, 
\emph{median}, and \emph{countdistinct}.  Each of these operators use a single pass
and require polylogarithmic space.  
%Each
%operator is discussed in more detail in Appendix \ref{appendix:operators}. 

Line \ref{example:filter} gives an example of a \emph{FILTER} statement.  The filter statement allows 
conditional down-selection of tuples.  \emph{Filtered}
is a partitioned stream of netflows, partitioned by destination IP, and then down-selected to only
allow netflows through where the average source total bytes is greater than 1000. 

%There are two other statement types not presented in Listing 1, 
%the TRANSFORM and COLLAPSE statements.  These will be discussed in the next section,
%with their use being demonstrated by a case study.
%
%It should be noted that the example program has the general form of a semi-streaming
%operation.  There are $N$ different streams, one for each destination IP.  For each of the $N$ streams
%we compute a $O(polylog \, n)$ spatial complexity streaming operation, 
%where $n$ is the window size.  Overall, the entire pipeline has spatial complexity $O(N \cdot polylog \, n)$,
%which closely resembles the definition of semi-streaming.  Generally speaking, most SAL
%programs follow a similar pattern.

Each SAL pipeline has two operating modes: training and testing.
In training, the pipeline is run against a finite set of data with labels,
and any features created by the pipeline will be appended per input tuple.  This
feature set along with the labels is used to train a classifier offline.
The testing phase then applies the pipeline to a live stream, transforming
each tuple into features, and then applies the trained classifier to the 
features to assign a label. 

%The language closest to SAL is Pig \cite{Olston:2008:PLN:1376616.1376726}.  Both are 
%imperative languages that describe the flow of data over a set of operations.
%However, given SAL's focus on streaming data instead of batch processing, the
%semantics of the two languages diverge.  The biggest difference is the
%\emph{imux} operation of the \emph{STREAM BY} statement.  While Pig processes 
%batches of tuples, SAL's main use case is to divide a never-ending stream of tuples into separate
%substreams, to each of which a pipeline of operations are applied.  Pig has a 
%\emph{FOREACH GENERATE} statement, but it's purpose is to transform a set of tuples
%into another set of tuples, whereas for SAL, the semantics are to create a feature
%for each substream of data.

\subsection{Case Study}
\label{section:caseStudy}
In this section, we take a look at one approach at creating a 
classifier for detecting botnets, namely
Disclosure \cite{Bilge:2012:DDB:2420950.2420969}.  
This will help elucidate how SAL can be used
to create succinct representations of machine learning 
pipelines that previously were developed in an
ad-hoc fashion.  Also, it will demonstrate what cannot 
be expressed by SAL.  Some of the 
features created by Disclosure require algorithms that 
do not comply with the desired constraints
of streaming and semi-streaming.  As such those 
features cannot be created with SAL.  However,
our intent with SAL is to down-select the stream of 
data to something more manageable for more
intensive study. 

A common architecture for botnets is to have a small set of command and control (C\&C)
servers that issue commands to a large number of infected machines, that then perform attacks
such as distributed denial-of-service, stealing data, spam 
\cite{SILVA2013378}, etc.
Disclosure focuses on identifying C\&C botnet servers.  
%They present a set of features based on flow sizes, how clients access servers, and temporal
%behaviors such as measuring diurnal fluctuations (or lack thereof).

The first part of the Disclosure pipeline identifies servers.  They define servers as an IP
address where the top two ports account for 90\% of the flows.  This can be expressed
in SAL with the \emph{TopK} operator, as in the following example:

%[caption=Finding Servers in Disclosure,label={lst:servers}]
\begin{lstlisting} 
VertsByDest = STREAM Netflows BY DestIp;
Top2 = FOREACH VertsByDest GENERATE topk(DestPort,10000,1000,2);
Servers = FILTER VertsByDest BY 
            top2.value(0) + top2.value(1) > 0.9;
\end{lstlisting}

As before in Listing \ref{lst:example}, we stream the netflows by destination IP.
Then with a \emph{FOREACH GENERATE} statement combined with the
\emph{topk} streaming operator, we calculate an estimate on the top two
ports that receive traffic for each destination IP.  We then follow that with a filter.
The \emph{value(n)} function returns the frequency of the $n^{th}$ most frequent
item (zero-based indexing).  

Once the servers have been determined, the authors of Disclosure hypothesized that
flow size distributions for C\&C servers are distinguishable from benign servers.
An example they give is that C\&C generally have a limited number of commands, and thus
flow sizes are limited to a small set of values.  On the other hand, benign servers will
generally have a much wider range of values.  To detect these difference between
C\&C servers and benign servers, they create three different types of features based on
flow size: statistical features, autocorrelation and unique flow sizes.
For the statistical features, they extract the mean and standard deviation for the size of incoming and
outgoing flows for each server.  This is easy to express in SAL as seen below.
%in Listing\ref{lst:statistical_features}.  
For each IP in the set of servers, we use the 
\emph{FOREACH GENERATE} statement combined with either the \emph{ave}
operator or \emph{var} operator.

%[caption=Statistical Features on Flow Sizes in Disclosure,label={lst:statistical_features}]
\begin{lstlisting}
FlowsizeSumIn = FOREACH Servers GENERATE ave(SrcTotalBytes);
FlowsizeSumOut = FOREACH Servers GENERATE ave(DestTotalBytes);
FlowsizeVarIn = FOREACH Servers GENERATE var(SrcTotalBytes);
FlowsizeVarOut = FOREACH Servers GENERATE var(DestTotalBytes);
\end{lstlisting}

Disclosure also generates features using autocorrelation on the flow
sizes.  The idea is that C\&C servers often have periodic behavior that 
an autocorrelation calculation would illuminate.
For each server, the sequence flow sizes can be thought of as a time series.  
They divide this signal up into 300 second
intervals and calculate the autocorrelation of the time series.  Unfortunately,
we are not aware of a streaming algorithm for calculating autocorrelation.
As such, we currently do not allow autocorrelation to be expressed within SAL; mixing
algorithms with vastly different spatial and temporal complexity requirements
would negate many of the benefits of the language.  However, if space is not
an issue, adding autocorrelation would be straightforward.

The final set of features based on flow-sizes involves finding the set of unique flow sizes.
The hypothesis here is that botnets have a limited set of messages, and so the number
of unique flow sizes will be smaller than a typical benign server.  To find an estimate on the
number of unique flow sizes, one can use the \emph{countdistinct} operator:

%[caption=Finding Number of Unique Flow Sizes, label={lst:unique_flow_sizes}]
\begin{lstlisting}
UniqueIn = FOREACH Servers GENERATE countdistinct(SrcTotalBytes);
UniqueOut = FOREACH Servers GENERATE countdistinct(DestTotalBytes);
\end{lstlisting} 

However, Disclosure goes a step further.  They create
an array with the counts for each unique element and then compute unspecified statistical features
on this array.  While this is not exactly expressible in SAL, and having an exact answer would break
our space constraints, one could use \emph{TopK} to obtain estimated counts for the most
frequent elements.

\begin{figure*}
\centering
\includegraphics[width=.75\linewidth]{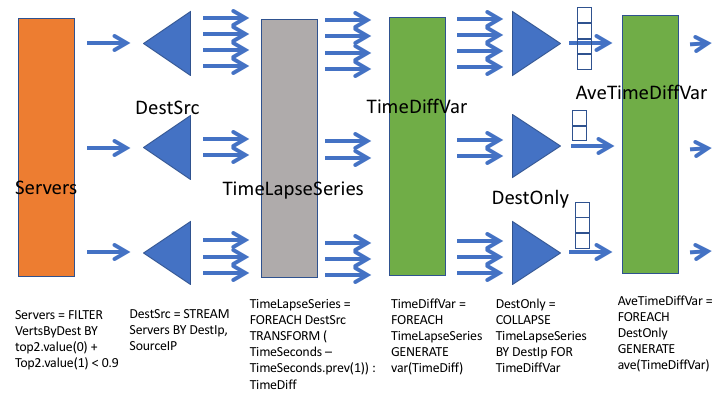}
\caption{This figure demonstrates the use of the \emph{TRANSFORM} and
\emph{COLLAPSE BY} statements to define parts of the Disclosure
pipeline.  }
\label{figure:slide6}
\end{figure*}

Besides features based on flow size, Disclosure also computes features on
client access patterns.  The hypothesis is that all the bots accessing a particular C\&C
server will exhibit very similar behavior, while the behavior of clients accessing benign
servers will not be so uniform.  Disclosure defines a time series for each server-client pair by
calcuating the inter-arrival times between consecutive connections.  For example, if we had
$n$ netflows that occurred at times $t_0$, $t_1$, ... $t_n$, then the series would be
$t_1 - t_0$, $t_2 - t_1$,  ...,$t_n - t_{n-1}$.  To specify this time series in SAL, one uses the 
\emph{TRANSFORM} statement.  The \emph{TRANSFORM}
statement allows one to transform from one tuple representation to another.

For this example, we need to transform from the original netflow tuple representation to a tuple
that has three values: the \emph{SrcIp}, \emph{DestIp}, and the inter-arrival time.
In the listing below, we first use the \emph{STREAM BY}
statement to further seprate the stream of netflows into source-destination IP pairs.
Then we follow that with the the \emph{TRANSFORM} statement that 
calculates the inter-arrival time.  Since \emph{SourceIp} and \emph{DestIp}
are the keys defined by the \emph{STREAM BY} statement, those values are
included by default as the first two values of the newly defined tuple.  
This part of the pipeline is represented in the left side of Figure \ref{figure:slide6}.

In the example below, we introduce the $prev(i)$ function.
The $prev(i)$ function returns the value of the related field $i$ items back
in time.  $TimeSeconds.prev(1)$ returns the value of $TimeSeconds$ 
in the tuple that occurred previous to the current tuple, thus giving us
the inter-arrival time.  The colon followed by \emph{TimeDiff} gives a label
to the tuple value and can be referred to in later SAL statements.
%(e.g. Listing \ref{lst:med_var}).

%[caption=Time Series Definition, label={lst:time_series}]
\begin{lstlisting}
DestSrc = STREAM Servers BY DestIp, SourceIp;
TimeLapseSeries = FOREACH DestSrc TRANSFORM  
               (TimeSeconds - TimeSeconds.prev(1)) : TimeDiff
\end{lstlisting}

Now that we have the time series expressed in SAL, we can then add the feature
extraction methods that Disclosure performs on the inter-arrival times.  Disclosure
calculates the minimum, maximum, median, and standard deviation.  The median
and standard deviation can be expressed in SAL below:
%as in Listing \ref{lst:med_var}.

%[caption=Feature Extraction on Time Series,label={lst:med_var}]
\begin{lstlisting}
TimeDiffVar = FOREACH TimeLapseSeries GENERATE var(TimeDiff);
TimeDiffMed = FOREACH TimeLapseSeries GENERATE median(TimeDiff);
\end{lstlisting}

However, maximum and minimum are not currently supported in SAL.  The reason
is that max and min require $O(n)$ space where $n$ is the size of the window
when computing over a sliding window \cite{Datar:2002:MSS:545381.545466}.  
When computing over the entire data stream, to compute the max/min
one can keep track of one number, but the sliding window adds complexity
as the max/min expires.  Perhaps some mixture between the two models,
over the entire stream and sliding windows, would be sufficient
for creating features.  Also, as with autocorrelation, if space is not an issue,
adding max and min is an easy extension.

%For example, you could use the scheme presented
%below to estimate the maximum value:
%
%\includegraphics[width=.75\linewidth]{figures/slides/Slide5}
%
%Here, we create a sequence of max values, $max_0$, $max_1$, ..., where each
%$max_i$ is calculated during the presentation $n$ consecutive items; 
%however, only two values need to be stored at
%any one time, $max_i$ and $max_{i+1}$.  When querying for the current max after $x$ items,
%then the index of the max value for item $x$ is 
%\begin{equation}
%f(x) = \begin{cases}
%0, & \text{if $x < \frac{n}{2}$.} \\
%\left \lfloor \frac{2}{n}  x -  1  \right \rfloor, & \text{otherwise.}    
%\end{cases}
%\end{equation}
%where $n$ is the size of the window.  This scheme would allow for constant space requirements
%and $O(n)$ processing time; however, it remains to be seen if max/min features extracted
%in this manner are useful in practical situations and we leave it as future work.

For the features derived from the time series to be applicable to classifying 
servers, we need to combine the features from all the clients. To do so, we
no longer use \emph{SourceIp} as one of the keys to separate the data flow 
by using the \emph{COLLAPSE BY} statement.
The \emph{BY} clause contains a list of keys that are kept.  Unspecified keys
are removed from the key set.  Below, 
%(see Listing \ref{lst:collapse}) 
\emph{DestIp} is kept while \emph{SourceIp} is not specified, meaning
that it is no longer a key to separate the data. 

\begin{lstlisting}
DestOnly = COLLAPSE TimeLapseSeries BY DestIp FOR TimeDiffVar, 
                                                  TimeDiffMed;
\end{lstlisting}

There are different possibilities for the semantics of the \emph{COLLAPSE BY} statement.
The one that we have implemented is the following: Let $k_{+}$ be the tuple elements
that are kept by \emph{COLLAPSE}, and let $k_{-}$ be the tuple elements that are no
longer used as a key.  For each tuple $t$ that appears in the stream $S$
of data, let $k_{+}(t)$ be the subtuple with only tuple elements from $k_{+}$,
and let $k_{-}(t)$ be the subtuple with only tuple elements from $k_{-}$.
Also, let $r(t)$ be the subtuple with remaining elements that are neither
in $k_{+}(t)$ nor in $k_{-}(t)$. 
Define $L$ to be the set of unique $k_{+}(t)$ for all $t \in S$, i.e.
%\begin{equation}
$L = \bigcup_{t \in S} k_{+}(t)$
%\end{equation}
For each $l \in L$, we define another set, $M_l$: 
%\begin{equation}
$M_l = \bigcup_{t \in S, k_{+}(t) =l} k_{-}(t)$
%\end{equation}
\emph{COLLAPSE BY} creates a mapping for each
set $M_l$, mapping the elements of $M_l$ to the most recently seen
$r(t)$ associated with each $m \in M_l$.  These mappings can then be
operated on by the
\emph{FOREACH GENERATE} statement.

Once we have collapsed back to \emph{DestIp}, we can then
calculate statistics across the set of clients for each server.  
The Disclosure paper does not specify which statistics are calculated, but
below we give some examples.

\begin{lstlisting}
AveTimeDiffVar = FOREACH DestOnly GENERATE ave(TimeDiffVar);
VarTimeDiffVar = FOREACH DestOnly GENERATE var(TimeDiffVar);
\end{lstlisting}

That concludes our exploration of how SAL can be used to express concepts
from a real pipeline defined in another paper.  While there are some operations
that are not supported, e.g. the autocorrelation features and max/min, most of the 
features could be expressed in SAL.  We believe SAL provides a succinct way
to express streaming machine learning pipelines.  In the next sections we discuss
how SAL is interpreted with a specific implementation.

\section{Implementation}
\label{section:implementation}
Here we discuss how SAL runs in parallel across a cluster.
We translate SAL with the Scala Parser Combinator Library \cite{scala-parser-combinator}.  
We express SAL's grammar and map those elements to C++ code that
utilizes a prototype parallel library that we wrote to execute SAL programs
called the \emph{Streaming Analytics Machine}, or SAM.
For the Disclosure pipeline, SAL uses 20 lines of code while SAM needs
520 lines.  For another pipeline we discuss in 
Section \ref{section:classifier_results}, SAL requires 34 lines while
SAM needs 520.  Overall, SAL uses 10-25 times fewer lines.  
%Table \ref{table:linecount} shows the difference
%between coding in SAM versus SAL.  SAL uses 10-40 times fewer lines.

%\begin{figure}
%\begin{minipage}{0.5\linewidth}
%\centering
%\begin{tabular}{| c | c | c |}
%\hline
%&C++ & SAL \\
%\hline
%Disclosure & 520 & 15\\
%\hline
%Results Pipeline & 482 & 32 \\
%\hline
%\end{tabular}
%\captionof{table}{Line count comparison.}
%\label{table:linecount}
%\end{minipage}
%\begin{minipage}{0.5\linewidth}
%\centering
%\includegraphics[width=.99\linewidth]{figures/architecture}
%\caption{Architecture of the system.}
%\label{figure:architecture}
%\end{minipage}
%\end{figure}

SAM is architected so that each node in the cluster 
receives tuple data.  Right now for the prototype, the only 
ingest method is a simple socket layer.  In maturing SAM, other options 
such as Kafka \cite{Garg:2013:AK:2588385} is an obvious alternative.
We then use ZeroMQ \cite{Akgul:2013:ZER:2523409} to distribute the tuples across the cluster.

For each tuple that a node receives, it performs a hash 
for each key specified in the \emph{PARTITION} statement, and sends the
tuple to the node assigned that key.  
For our experiments we partition on both source IP and dest IP, meaning
that for each netflow a node receives over the socket layer, it sends the same 
netflow twice over ZeroMQ (if the netflow is not kept locally).
Many different messaging styles can be expressed by ZeroMQ.
We make use of the push/pull paradigm of ZeroMQ.  Each node creates $n-1$
push sockets and $n-1$ pull sockets, where $n$ is the size of the cluster.
%Continuing with our example of partitioning by IP, 
%each time a netflow arrives, a node hashes the source and destination IP addresses
%and sends the netflow over the appropriate pull socket.  On the receiving end, each node uses
%zmq\_poll to read from the $n$ pull sockets.  Another option was to use a publish/subscribe
%paradigm; however, the publish/subscribe messaging style allows for dropped items once
%the high-water mark is reached.  Push sockets block until the pull socket queue has room for new items.
%In this work we wanted to ensure that none of the data is lost.  However, in other settings
%it may be acceptable for some loss to occur.  Streaming operations are inherently estimations,
%so some degree of loss may prove inconsequential to the final results.

Conceptually, many of the statements and operators map to either consumers and/or producers,
which we implement with C++ classes. 
The prototype implementation reads netflow data from a socket, so the class ReadSocket
reads from the socket and is considered the original source of the stream.  A ZeroMQPushPull
instance, acting as a consumer, takes the data from the ReadSocket and distributes the netflow
data across the cluster using the push sockets.  The same ZeroMQPushPull instance uses
the pull sockets to collect the netflows destined for it.  It then collects those netflows in a queue and
once full, calls a parallel \emph{feed} method.  The feed method provides the contents of the queue
to all registered consumers in parallel.
%
%\begin{table}
%\footnotesize
%\centering
%\begin{tabular}{| c | c | c | c | c |}
%\hline
%C++ Name & Con- & Pro- & Feature & Maps To \\
%& sumer &ducer & Creator & \\
%\hline
%ReadSocket & & x & & VastStream \\
%\hline
%ZeroMQPushPull & x & x & & N/A \\
%\hline
%Filter & x & x & & FILTER \\
%\hline
%Transform & x & x & & TRANSFORM \\
%\hline
%CollapsedConsumer & x & & x & COLLAPSE \\
%\hline
%Project & x & & & COLLAPSE \\
%\hline
%EHSum & x & & x & sum \\
%\hline
%EHAve & x & & x & ave \\
%\hline
%EHVariance & x & & x & var \\
%\hline
%TopK & x & & x & topk \\
%\hline
%\end{tabular}
%\caption{A listing of the major classes in SAM and how they map to language features in SAL.
%We also make the distinction between consumers, producers, and feature creators.  Producers
%all share a parallel \emph{feed} method that sends the data to registered consumers, which is 
%how parallelization is achieved.  Each of the consumers receive data from producers.  If a consumer
%is also a feature creator, it performs a computation on the received data, generating a feature,
%which is then added to a thread-safe feature map.}
%\label{table:classes}
%\end{table}

Often consumers generate features.  Each node creates a thread-safe feature map
to collect features that are generated.  The function signatures is as follows:

\begin{lstlisting}
updateInsert(string key, string featureName, Feature f)
\end{lstlisting}
The \emph{key} is generated by the key fields specified in the
\emph{STREAM BY} statement.  The \emph{featureName}
comes from the identifier specified in the \emph{FOREACH GENERATE}
statement.  For example,
in the below SAL snippet, the key is created by using a string hash
function on the concatenation of the source IP and destination IP.
The identifier is \emph{Feature1} as specified in the 
\emph{FOREACH GENERATE} statement.
The scheme for ensuring thread-safety in the feature map comes from Goodman et al.  \cite{Goodman:2011:SHS:2063384.2063439}.
\begin{lstlisting}
DestSrc = STREAM Netflows BY SourceIp, DestIp;
Feature1 = FOREACH DestSrc GENERATE ave(SrcTotalBytes);
\end{lstlisting}

The \emph{Project} class provides the functionality of
the \emph{COLLPASE} statement.  The \emph{Project} class creates features similar
to the Feature Creator classes, but they cannot be accessed directly through the API.
The features are \emph{Map} features, in other words the sets $M_l$ using the terminology from
Section \ref{section:caseStudy}.  These \emph{Map} features are added to the same
feature map used for all other generated features.  These \emph{Map} features
can then be used by the \emph{CollapsedConsumer} class, which can be specified
to calculate statistics on the map for each kept key in the stream.

\section{Classifier Results}
\label{section:classifier_results}
To validate the value of SAL in expressing pipelines and in filtering out benign data, we used a simple
program where we separate the netflows two ways, by destination IP and by source IP.  Then
we create features based on the $ave$ and $var$ streaming operators for each of the fields: 
$SrcTotalBytes$, $DestTotalBytes$, $DurationSeconds$, 
$SrcPayloadBytes$, $DestPayloadBytes$, $SrcPacketCount$
and $DestPacketCount$.  This results in 28 total features.

We apply this pipeline to CTU-13 \cite{GARCIA2014100}.
CTU-13 has nice characteristics, including:
1) Real botnet attacks: Virtual machines were created and infected.
2) Real background traffic: Traffic from their university router was captured at the same time 
as the botnet traffic.  The botnet traffic was bridged into the university network.
3) A variety of protocols and behaviors: The scenarios cover a range of protocols that were used by the malware such as IRC, P2P, and HTTP.  Also some scenarios sent spam, others performed
click-fraud, port scans, DDoS attacks, or Fast-Flux.
4) A variety of bots: The 13 scenarios use 7 different bots.

SAL program inherently encodes historical information.
As such we can't treat each netflow 
independently and thus can't create random subsets of the data as is 
usual in a cross-validation approach.  We instead
split each scenario into two parts.  We want to keep the number of 
malicious netflows about the same in each part (we need enough 
examples to train on), so we find the point in the scenario 
timewise where the malicious examples are balanced.  
Namely we have two sets, $P_1$ and $P_2$, where 
$\forall n_1 \in P_1, \forall n_2 \in P_2,
 TimeSeconds(n_1) < TimeSeconds(n_2) $ 
 and $| Malicious(P_1) | \approx |Malicious(P_2)| $,
 where $TimeSeconds$ returns the time in seconds of the netflow
 and $Malicious$ returns a subset of the provided set of all the malicious
 netflows in the provided set.
With each scenario split into two parts, we train on the first part and test on the
second part.  Then we switch: train on the second and test on the first.
We make use of a Random Forest Classifier as implemented in 
scikit-learn \cite{scikit-learn}.

After generating the 28 features, we performed
a greedy search over them to down-select to the most important ones.
We added features one at a time
until no improvement is found in the average AUC over the 13 scenarios.  
The following eight features provided the best performance across the
13 scenarios.  They are listed in the order they were added using the greedy approach 
above: 1) Average \emph{DestPayloadBytes},
2) Variance \emph{DestPayloadBytes},
3) Average \emph{DestPacketCount},
4) Variance \emph{DestPacketCount},
5) Average \emph{SrcPayloadBytes},
6) Average \emph{SrcPacketCount},
7) Average \emph{DestTotalBytes}, and
8) Variance \emph{SrcTotalBytes}.

Using the above eight features, Figure \ref{figure:ctu} shows the AUC of the 
ROC for each of the 13 scenarios.  
In four scenarios, 1, 3, 6, and 11, training on either half was sufficient for the
other half, with AUCs between 0.926 and 0.998.  Some scenarios, 
namely 2, 5, 8, 12, and 13, the first half was sufficient to obtain AUCs between 0.922
and 0.992 on the second half, but the reverse was not true.  For scenario 10,
training on the second half was predictive of the first half, but not the other way.
Scenario 9 had AUCs of 0.820 and 0.887, which is decent, but lackluster
compared to the other scenarios.
The classifier had issues with scenarios 4 and 7.  
Scenario 7 did not perform well, probably because there were only 63 malicious netflows.  
We are not sure why the classifier struggled with scenario 4.

%TODO: Comparison to other wor

%TODO: Add about filtering data

\section{Scaling}
\label{section:scaling_results}
For our scaling experiments, we use Cloudlab \cite{RicciEide:login14}, 
a set of clusters distributed
across three sites, Utah, Wisconsin, and South Carolina, where researchers can provision a set of 
nodes to their specifications.  We created an image where our code, SAM, was deployed
and working, and then replicated that image to a cluster size of our choice. 
In particular we make use of the Clemson system in South Carolina.  The Clemson system has 16 cores per node, 10 Gb/s Ethernet and 256 GB of memory.  We were able to allocate a cluster
with 64 nodes.

\begin{figure}[t]
\centering
\begin{minipage}{.45\textwidth}
\centering
\includegraphics[width=.99\linewidth]{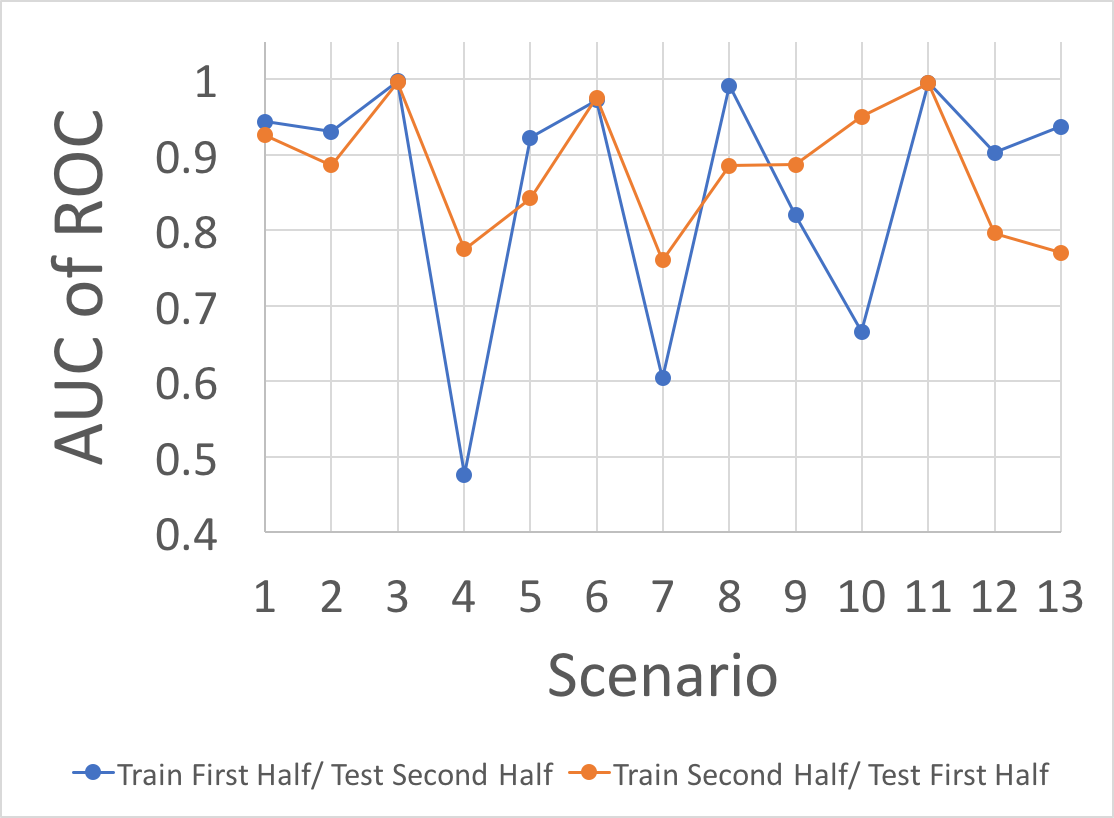}
\captionof{figure}{AUC of the ROC
%This graphic shows the AUC of the ROC curve
%for each of the CTU scenarios.  We first train on the first half
%of the data and test on the second half, and then switch.     
}
\label{figure:ctu}
\end{minipage}
\begin{minipage}{.45\textwidth}
  \centering
  \includegraphics[width=.99\linewidth]{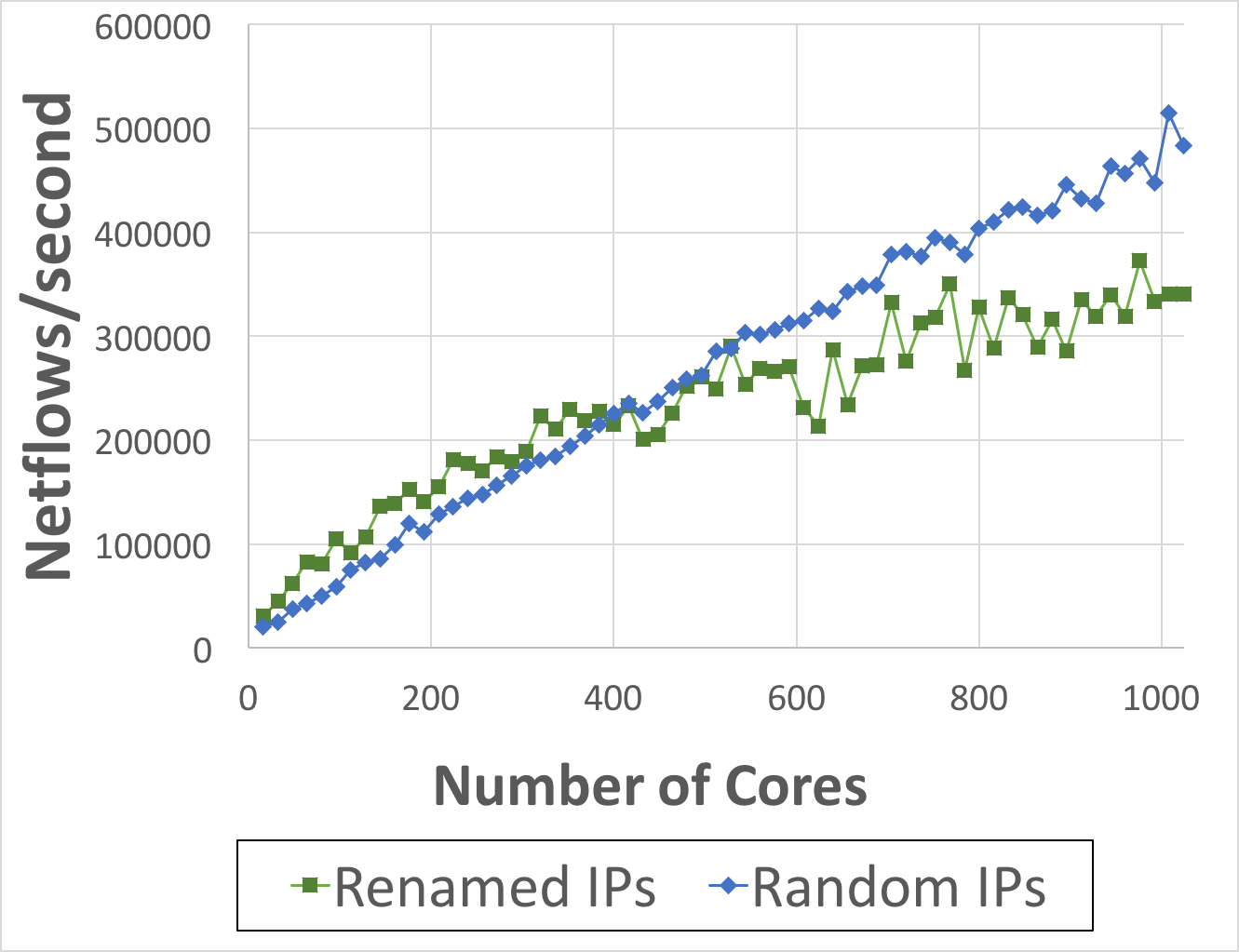}
  \captionof{figure}{Weak Scaling Results
  %: For each run with $n$
  %nodes, a total of $n$ million netflows were run through the system with
  %a million contiguous netflows per node randomly selected from the CTU dataset. 
  %There were two types of runs, with renamed IPs (to create the illusion of a larger network)
  %or with randomized IPs.  For the renamed IPs, there continues to be improved
  %throughput until 61 nodes/ 976 cores.  For the randomized IPs, the scaling continues
  %through the total 64 nodes / 1024 cores.
 }
  \label{figure:scaling}
\end{minipage}%
\begin{minipage}{.45\textwidth}
  \centering
  \includegraphics[width=.99\linewidth]{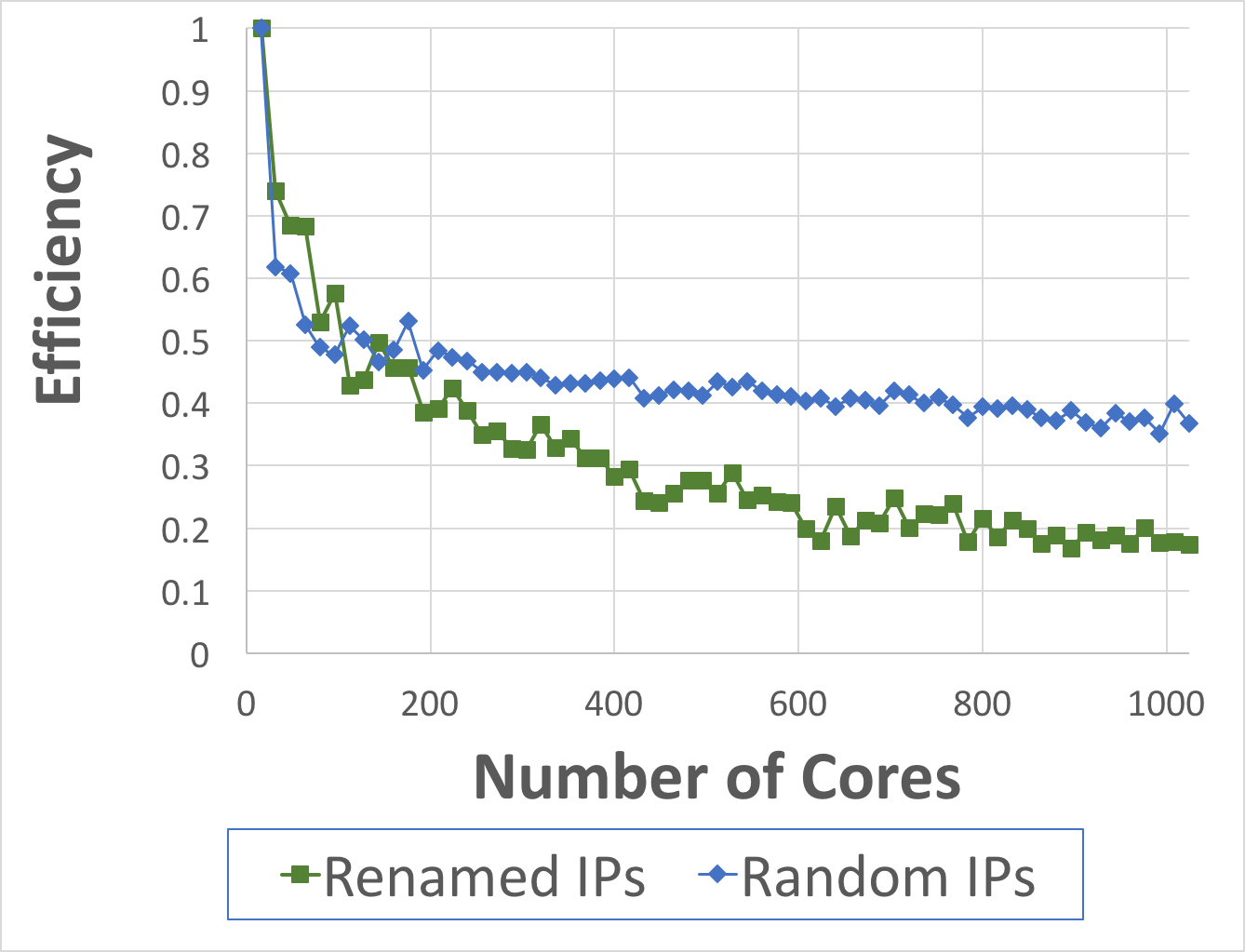}
  \captionof{figure}{Weak Scaling Efficiency.  
  %The efficiency continues to
  %degrade for the renamed IPs, but for the randomized IPs, the efficiency hovers a little above
  %0.5.  This is largely due to the fact that the randomized IPs almost perfectly load-balances
  %the work, while the renamed IPs encounters load imbalances because of the power-law
  %nature of the data.
  }
  \label{figure:efficiency}
\end{minipage}
\end{figure}

Figure \ref{figure:scaling} shows the weak scaling results.  Weak scaling examines the 
solution time where the problem size
scales with the number of nodes, i.e. there is a fixed problem size per node.
For our experiments, each node was fed one million netflows.
Thus, for $n$ nodes, the total problem size is $n$ million netflows.
Each node had available the entire CTU dataset concatenated into one file.  Then we randomly
selected for each node a contiguous chunk of one million netflows.    
For each chunk, we renamed the IP addresses
to simulate a larger network instead of replaying the same IP addresses, just from different time
frames.  For comparison we also ran another round with completely random IP addresses, such
that an IP address had very little chance of being in multiple netflows.  This helped us determine
if scaling issues on realistic data were from load balancing problems or some other issue.
Each point
is the average of three runs.

In Figure \ref{figure:scaling} we see the renamed 
IP set of runs peaks out at 61 nodes or 976 cores
where we obtain a throughput of 373,000 netflows per second or 32.2 billion
 per day.
 For the randomized IP set of runs, the scaling is noticeably steeper, reaching a peak 
 throughput of 515,000 netflows per second (44.5 billion per day) with 63 nodes.
 Figure \ref{figure:efficiency} takes a look
 at the weak scaling efficiency.  This is defined as $E(n) = T_1/ T_n$, where $T_i$ is the time
 taken by a run with $i$ nodes.  Efficiency degrades quicker for the renamed IP set of runs 
 while the randomized IP runs hover close to 0.4.
 We believe the difference is due to load balance issues.
 For the randomized IP
 set of runs, the work is completely balanced
 between all 64 nodes.  The renamed IP runs are more realistic, akin to a power law distribution, where
 a small set of nodes account for most of the traffic.  In this situation, it becomes difficult to 
 partition the work evenly across all the nodes.  

As far as we know, we are the first to show scalable 
distributed network analysis on a cluster of size 64 nodes.
The most direct comparison in terms of scaling is Bumgardner and
 Marek \cite{Bumgardner:2014:SHS:2568088.2568103}.  In this work, they 
 funnel netflows through a 10 node cluster running Storm \cite{storm}.
 They call their approach a hybrid stream/batch system because it uses 
 Storm to stream netflow data into a batch system, Hadoop \cite{hadoop}, for
 analysis.  The stream portion is what is most similar to our work.  Over the
 Storm pipeline, they achieve a rate of 234,000 netflows per second, or about 23,400 
 netflows per second, per
 node.  Our pipeline with ten nodes achieved a rate of 13,900 netflows per second,
 per node.  However, their pipeline is much simpler.  They do not partition the netflows
 by IP and calculate features.  The only processing they undergo during streaming
 is adding subnet information to the netflows, which does not require partitioning across 
 the cluster.  
 
 %There is other work which performs distributed network analysis, but their focus is
 %on packet-level data
 %\cite{Lee:2012:TSI:2427036.2427038,7535295,6906761}.  
 %As such it is difficult to compare as their metrics are in terms
 %of GBs/second, and of course the nature of packet analysis is significantly different
 %than netflow analysis.
 
 In terms of botnet identification, Botfinder \cite{Tegeler:2012:BFB:2413176.2413217}
 and Disclosure \cite{Bilge:2012:DDB:2420950.2420969} report
 single node batch performance numbers.  Our approach on a single 16-core node achieved a rate
 of 30,500 netflows per second.  For Botfinder, they extract 5 features
 on a 12 core Intel Core i7 chip, achieving a rate of 46,300 netflows per second.
 Disclosure generates greater than nine features (the text is ambiguous) on
 a 16 core Intel Xeon CPU E5630.  They specify that they run the feature extraction
 for one day's worth of data
 in 10 hours and 53 minutes, but they do not clearly state if that is on both
 data sets they chose to evaluate or just one.  
 If it is both, the rate is roughly 40,000 netflows per second.  

\section{Related Work}
\label{section:related}

There are many frameworks that provide streaming APIs.  
Prominent among them are Apache Storm \cite{storm}, 
Apache Spark \cite{spark}, Apache Flink \cite{ApacheFlink},
and Apache Heron \cite{heron}.  Each has different advantages 
and short comings.  Many perform checkpointing to allow for replay 
in case of failure.  However, the price
of checkpointing may be too significant a cost to justify for our application.  
Our streaming calculations are by definition approximations, so some lost
data may be acceptable.  Zhang et al. \cite{wukong-2017} report success
in adapting Storm as the backend of a streaming C-SPARQL engine 
\cite  {c-sparql-2010}, while Spark has significantly longer latencies.
In our own experiments, we found Spark Streaming to have trouble scaling to the
node counts we used.  We have also experimented with Flink, which has
a wealth of implemented streaming concepts that are a natural fit for
SAL.  So far we've found mixed results SAM vs Flink, 
but that is outside the current scope of this paper will be reported in future work.
In any case, our main contribution in this paper is the domain specific language
for expressing cyber queries.  The underlying implementation can be changed
and adapted as technology evolves.

In terms of domain specific languages, there are several graph-related DSL's
that allow for vertex-centric computations, similar to our feature calculations
on a per vertex basis.
DSL's like Green-Marl \cite{green-marl-2012} and Ligra \cite{ligra-2013}
provide succint ways to express graph computations, but they are 
limited to shared-memory infrastructures.  Other approaches
like Gluon \cite{gluon-2018} provide a mechanism for
converting shared-memory approaches to distributed settings.
Regardless of whether these DLS approaches are computationally distributable, 
a fundamental difference between this set
of work and our own is the streaming aspect of our aproach and domain.
The algorithmic solutions, partitioning,
operation scheduling, and data processing of these graph DSL's
rely upon the assumption of static data.
Also, these approaches do not have a way of expressing machine
learning pipelines.

%Yet another Apache project that relates to the current work is Kafka \cite{Garg:2013:AK:2588385}.
%Kafka is a distributed streaming platform that allows you to build reliable streaming pipelines.
%It has producer and consumer APIs, similar to concepts we discuss in 
%Section \ref{section:implementation}.  Along with a Streams API that allows you to transform
%streams, and a Connector API, that allows you to connect to existing systems.  The same 
%argument applies, that Kafka could be used as the basis of an implementation, but given
%the domain we are targeting, a custom implementation will almost certainly avoid
%unneeded overheads.

%S4

\section{Conclusions}
\label{section:conclusions}

We have presented a new domain specific language, the 
Streaming Analytics Language, or SAL, that is designed to
easily express analytical pipelines on streaming data.  We specifically examined the case of
cyber data and showed how it can extract features from netflow data which can then be used to
train a classifier using labeled data.  Using the CTU-13 dataset as an example, we were able to
train classifiers on streaming features 
that on average obtained an AUC of 0.87.  In an operational setting, this could be used
to greatly reduce the amount of traffic needing to be analyzed.  SAL can be used as a first pass over
the data using space and temporally-efficient streaming algorithms.  After down-selecting, 
more expensive algorithms can be applied to the remaining data.

In addition to SAL, we also presented the results of our scalable interpreter of SAL, which we call
the Streaming Analytics Machine, or SAM.  On real data, we are able to scale to 61 nodes and 976 cores,
obtaining a throughput of 373,000 netflows per second or 32.2 billion
 per day.  On completely load-balanced data, we obtain greater efficiency out to 64 nodes than
 the real data, indicating that SAM could be
 improved with a more intelligent partitioning strategy.  In the end, we have an easy to use
 domain specific language and a scalable implementation to back it that has good accuracy on
 the target problem.

\bibliographystyle{splncs04}
\bibliography{bib,graph_dsls}

\end{document}